\documentclass[twocolumn]{revtex4}
\usepackage{graphicx}
\usepackage{bm}
\usepackage{color}
\usepackage{amsmath}
\usepackage{natbib}

\begin{document}

\title{Separated spin-up and spin-down evolution of degenerated electrons in two dimensional systems: Dispersion of longitudinal collective excitations in plane and nanotube geometry}

\author{Pavel A. Andreev}
\email{andreevpa@physics.msu.ru}
\author{L. S. Kuz'menkov}%
\email{lsk@phys.msu.ru}
\affiliation{Faculty of physics, Lomonosov Moscow State University, Moscow, Russian Federation.}

 \date{\today}

\begin{abstract}
Applying the separated spin evolution quantum hydrodynamics to two-dimensional electron gas in plane samples and nanotubes located in external magnetic fields we find new kind of wave in electron gas, which is called the spin-electron acoustic wave. Separate spin-up electrons and spin-down electrons evolution reveals in replacement of the Langmuir wave by the couple of hybrid waves. One of two hybrid waves is the modified Langmuir wave. Another hybrid wave is the spin-electron acoustic wave. We study dispersion of these waves in two dimensional structures of electrons. We also consider dependence of dispersion properties on spin polarisation of electrons in external magnetic field.
\end{abstract}

\pacs{52.30.Ex, 52.35.Dm, 73.22.Lp}% PACS, the Physics and Astronomy
                             % Classification Scheme.
\keywords{quantum plasmas, quantum hydrodynamics, 2DEG, nanotubes, spin evolution}
%Use showkeys class option if keyword

\maketitle

%52.30.Ex	Two-fluid and multi-fluid plasmas
%52.35.Dm	Sound waves

%52.27.Ep	Electron-positron plasmas

%%%%%%%%%%TEXT

The Langmuir wave is a fundamental process in multy electron three dimensional and low dimensional  systems creating a background for many applications. Field of plasminics one of great examples of these applications. However, considering separate evolution of spin-up and spin-down electrons moving in an external magnetic field we find that collection of electrons reveals new collective excitation: the spin-electron acoustic wave \cite{Andreev spin-up and spin-down 1405}. This wave has been recently predicted and studied for three dimensional electron gas \cite{Andreev spin-up and spin-down 1405}. Properties of the spin-electron acoustic wave and modification of properties of the Langmuir wave in two dimensional electron gases (2DEGs) having plane and cylindric (nanotube) geometry is under theoretical consideration in this paper.

Carbone-nanotubes are most famous nanotubes, since they have been used in different application. However there are other types of nanotubes: gold \cite{Zhou PRB 07}, \cite{Manrique PRB 10}, \cite{Kohl PRB 11}, silicon-based \cite{Guzman-Verri PRB 07}, platinum \cite{Konar PRB 08} nanotubes, and MgO and $\textrm{Fe}_{3}\textrm{O}_{4}$-based nanostructures \cite{Mejia-Lopez PRB 14}.

Hydrodynamic description of nanotubes can be found in Refs. \cite{Longe PRB 93}-\cite{Khan JAP 14}. $\pi$ electron--hole plasma in single-walled metallic carbon nanotubes is considered in Ref. \cite{Moradi PLA 08} in terms of two fluids hydrodynamics.

Carbone-nanotubes contain two types of electrons belonging to $\pi$ and $\sigma$ orbitals. They concentrations are different $ n_{0\sigma}=3n_{0\pi}$. That reveals in splitting of the Langmuir wave on two waves \cite{Moradi PRB 07}, \cite{Dmitrovic PRB 08}. This effect is described by the two liquid hydrodynamics, where electrons on $\pi$ and $\sigma$ orbitals are considered as different interacting species \cite{Moradi PRB 07}, \cite{Dmitrovic PRB 08}.

In opposite to electron gas, electrons in graphite and carbone nanotubes are bounded to atoms. Three of four valence electrons are in strong $\sigma$ bonds, and one electron occupies $\pi$ orbital. At hybridisation of orbitals electrons occupy all for states with different modules and projection of the orbital momentum (one s state and three p states) having same spin projection. Thus valence electrons in carbone nanotubes are fully polarised. Since electrons of carbone nanotubes are fully polarised, we do not expect the spin-electron acoustic wave in carbone nanotubes.

Influence of spin on nanotubes properties have been recently discussed in Ref. \cite{Palyi PRL 12}.
Energy loss of a plasmon in a disorder-free carbon nanotube and plasmon decays into neutral bosonic excitations of the electron liquid were considered in Ref. \cite{Chen PRB 10}.

Main tool of our research is the separated spin evolution quantum hydrodynamics (SSE-QHD) derived in Ref. \cite{Andreev spin-up and spin-down 1405}. The SSE-QHD is a generalisation of the spin-1/2 quantum hydrodynamics \cite{Takabayasi PTP 55 a}, \cite{MaksimovTMP 2001}, \cite{Marklund PRL07}, \cite{Andreev RPJ 07}.
Time evolution of the two dimensional particle densities (concentrations) $n_{u}$ and $n_{d}$ gives the continuity equations
\begin{equation}\label{SUSD2D cont eq electrons spin s}
\partial_{t}n_{s}+\nabla(n_{s}\textbf{v}_{s})=(-1)^{i_{s}}T_{z}, \end{equation}
where $s=\{u=\uparrow, d=\downarrow\}$, $[n_{s}]=$cm$^{-2}$, $T_{z}=\frac{\gamma}{\hbar}(B_{x}S_{y}-B_{y}S_{x})$ is the z-projection of torque presented in Cartesian coordinates, $i_{s}$: $i_{u}=2$, $i_{d}=1$, with the spin density projections $S_{x}$ and $S_{y}$, each of them is a mix of $\psi_{\uparrow}$ and $\psi_{\downarrow}$, which are components of the wave spinor. Explicit form of $S_{x}$ and $S_{y}$  appear as $S_{x}=\psi^{*}\sigma_{x}\psi=\psi_{d}^{*}\psi_{u}+\psi_{u}^{*}\psi_{d}=2a_{u}a_{d}\cos\Delta \phi$, $S_{y}=\psi^{*}\sigma_{y}\psi=\imath(\psi_{d}^{*}\psi_{u}-\psi_{u}^{*}\psi_{d})=-2a_{u}a_{d}\sin\Delta \phi$, where $\Delta \phi=\phi_{u}-\phi_{d}$. These quantities do not related to different species of electrons having different spin direction. $S_{x}$ and $S_{y}$ describe simultaneous evolution of both species. Hence $S_{x}$ and $S_{y}$ do not wear subindexes $u$ and $d$. We will apply the QHD equations for plane and cylindric geometries, but we have presented the torque in the Cartesian coordinates only, with no representation in the cylindrical coordinates. We have it done since it is a non-linear term. It does nor affect linear properties of electron gas, which are considered in this paper. We present some non-linear terms in the Euler equation, below, in the same manner.
We note that $\textbf{v}=\{v_{x},v_{y}\}$ and $n=n(x,y)$, $\textbf{v}=\textbf{v}(x,y)$ for plane-like 2DEG, \emph{and} $\textbf{v}=\{v_{\varphi},v_{z}\}$, and $n=n(\varphi,z)$, $\textbf{v}=\textbf{v}(\varphi,z)$ for nanotubes.

The time evolution of the particle currents for each projection of spin $\textbf{j}_{u}=n_{u}\textbf{v}_{u}$ and $\textbf{j}_{d}=n_{d}\textbf{v}_{d}$ gives Euler equations
$$mn_{s}(\partial_{t}+\textbf{v}_{s}\nabla)\textbf{v}_{s}+\nabla p_{s}-\frac{\hbar^{2}}{2m}n_{s}\nabla\Biggl(\frac{\triangle \sqrt{n_{s}}}{\sqrt{n_{s}}}\Biggr)$$
$$=q_{e}n_{s}\biggl(\textbf{E}+\frac{1}{c}[\textbf{v}_{s},\textbf{B}]\biggr)+(-1)^{i_{s}}\frac{\gamma_{e}}{m}n_{s}\nabla B_{z}$$
\begin{equation}\label{SUSD2D Euler eq electrons spin UP} +\frac{\gamma_{e}}{2m}(S_{x}\nabla B_{x}+S_{y}\nabla B_{y})+(-1)^{i_{s}}(\widetilde{\textbf{T}}_{z}-m\textbf{v}_{s}T_{z}),\end{equation}
with $\widetilde{\textbf{T}}_{z}=\frac{\gamma_{e}}{\hbar}(\textbf{J}_{(M)x}B_{y}-\textbf{J}_{(M)y}B_{x})$, which is the torque current, where
\begin{equation}\label{SUSD2D Spin current x} \textbf{J}_{(M)x}=\frac{1}{2}(\textbf{v}_{u}+\textbf{v}_{d})S_{x}-\frac{\hbar}{4m} \biggl(\frac{\nabla n_{u}}{n_{u}}+\frac{\nabla n_{d}}{n_{d}}\biggr)S_{y}, \end{equation}
and
\begin{equation}\label{SUSD2D Spin current y} \textbf{J}_{(M)y}= \frac{1}{2}(\textbf{v}_{u}+\textbf{v}_{d})S_{y}+\frac{\hbar}{4m}\biggl(\frac{\nabla n_{u}}{n_{u}}+\frac{\nabla n_{d}}{n_{d}}\biggr)S_{x}, \end{equation}
where $q_{e}=-e$, $\gamma_{e}=-g\frac{e\hbar}{2mc}$ is the gyromagnetic ratio for electrons, and $g=1+\alpha/(2\pi)=1.00116$, where $\alpha=1/137$ is the fine structure constant, so we include the anomalous magnetic moment of electrons. $\textbf{J}_{(M)x}$ and $\textbf{J}_{(M)y}$ are elements of the spin current tensor $J^{\alpha\beta}$. $\textbf{J}_{(M)x}$, $\textbf{J}_{(M)y}$, $\widetilde{\textbf{T}}_{z}$, $T_{z}$ are non-linear terms, and they do not give contribution in spectrum.
$p$ is the thermal pressure, or the Fermi pressure for degenerate fermions. Term proportional to the square of the Plank constant is the quantum Bohm potential.

The right-hand side of Euler equation present force fields of interaction. The first groups of terms in the right-hand side are the Lorentz forces. The second terms describe action of the z-projection of magnetic field on the magnetic moments (spins) of particles. Dependence on spin projection reveals in different signs before these terms. The third groups of terms in Euler equations contain a part of well-known force field $\textbf{F}_{S}=M^{\beta}\nabla B^{\beta}$ describing action of the magnetic field on magnetic moments \cite{Takabayasi PTP 55 a}, \cite{MaksimovTMP 2001}. Part of this force field has been presented by previous terms $\textbf{F}_{S(z)}=\pm\gamma_{e}n_{u,d}\nabla B_{z}$. The second part of the force field $\textbf{F}_{S(x,y)}=\gamma_{e}(S_{x}\nabla B_{x}+S_{y}\nabla B_{y})$. The half of this force field enters each of the Euler equations. The last groups of terms is related to nonconservation of particle number with different spin-projection. This nonconservation gives extra mechanism for change of the momentum density revealing in the extra force fields.

At transition to the cylindric coordinates the inertia forces appear in hydrodynamic equations. These forces consist of three parts: the convective part (containing the velocity field $\textbf{v}$), the thermal part (containing pressure $p$), and the quantum part (proportional to the square of the Planck constant $\hbar^{2}$). If the thermal pressure is isotropic, as we consider in this paper, the thermal pressure appears in the Euler equation in traditional form $\nabla p$, with no extra terms. The convective part of the inertia forces gives no contribution in linear exitations on cylindric surface. The quantum part of the inertia forces together with the quantum part of the momentum flux tensor can be presented as the third term in equation (\ref{SUSD2D Euler eq electrons spin UP}). This form coincides with the traditional form of the quantum Bohm potential in Cartesian coordinates.

Spin evolution itself leads to new collective excitations \cite{Andreev VestnMSU 2007}-\cite{Shukla RMP 11}, which are, some times, called the spin-plasma waves, but we do not consider them here.

We do not include influence of spin evolution on longitudinal waves. Thus we do not present equations for these quantities. These equations can be found in Ref. \cite{Andreev spin-up and spin-down 1405}.

A model aimed to describe separated evolution of spin-up and spin-down electrons was considered in Ref. \cite{Brodin PRL 10 SPF}, but electrons with different spin projections are  described there in the same manner as electrons with no spin separation. However it contradicts to the model directly derived from the Pauli equation, independently, for spin-up and spin-down electrons \cite{Andreev spin-up and spin-down 1405}.

Electric and magnetic fields in the Euler equation (\ref{SUSD2D Euler eq electrons spin UP}) have the following explicit relation with sources of fields
\begin{equation}\label{SUSD2D El field} \textbf{E}=-q_{e}\nabla\int \frac{n_{u}+n_{d}-n_{0}}{\mid \textbf{r}-\textbf{r}'\mid}d\textbf{r}',\end{equation}
and
\begin{equation}\label{SUSD2D} \textbf{B}=\int  [(\textbf{M}\nabla)\nabla-\textbf{M}\triangle]\frac{1}{\mid \textbf{r}-\textbf{r}'\mid}d\textbf{r}',\end{equation}
where $d\textbf{r}'$ is a differential of two dimensional surface: $d\textbf{r}'=dxdy$ for planes, and  $d\textbf{r}'=Rd\varphi dz$ for cylinders. $n_{0}$ in equation (\ref{SUSD2D El field}) presents motionless ions. $\textbf{B}=\textbf{B}(x,y)$, or $\textbf{B}=\textbf{B}(\varphi,z)$, and $\textbf{B}=\{ B_{x},B_{y}, B_{z}\}$ or $\textbf{B}=\{ B_{r},B_{\varphi}, B_{z}\}$, structure of $\textbf{M}$ is similar to $\textbf{B}$.

Equation of state for the pressure of spin-up $p_{\uparrow}$ and spin-down $p_{\downarrow}$ degenerate electrons, appears as
$p_{s}=\pi\frac{\hbar^{2}}{m}n_{s}^{2}$.
Pressure of spin-up electrons and spin-down electrons are different due to the external magnetic field, which changes an equilibrium concentration of each species $n_{0\uparrow}\neq n_{0\downarrow}$. In pressure $p_{s}$ we have included that only one particle with a chosen spin direction can occupy one quantum state. As a consequence coefficient in the equation of state two times bigger than in the 2D Fermi pressure.

Equilibrium condition is described by the non-zero concentrations $n_{0\uparrow}$, $n_{0\downarrow}$, $n_{0}=n_{0\uparrow}+n_{0\downarrow}$, and the external magnetic field $\textbf{B}_{ext}=B_{0}\textbf{e}_{z}$. Other quantities equal to zero $\textbf{v}_{0\uparrow}=\textbf{v}_{0\downarrow}=0$, $\textbf{E}_{0}=0$, $S_{0x}=S_{0y}=0$. If we consider plane-like 2DEGs we place it in the plane $z=0$, perpendicular to the external magnetic field. Hence waves propagate perpendicular to the external magnetic field. Perturbations of physical quantities are presented as $\delta f=F(\omega, k_{x}, k_{y})e^{-\imath\omega t+\imath k_{x}x+\imath k_{y}y}$ and $k^{2}=k_{x}^{2}+k_{y}^{2}$, with $\delta f=\{\delta n_{u},\delta n_{d}, \delta \textbf{v}_{u}, \delta \textbf{v}_{d}\}$. If we consider nanotubes we place them parallel to the external magnetic field.
We present perturbations in the following form
%\begin{equation}\label{SUSD2D perturbations}\end{equation}
$\delta f=\int\sum_{l=0}^{\infty}F_{l}(k,\omega)e^{-\imath\omega t+\imath kz+\imath l\varphi} dk d\omega$.
Representation of perturbations via exponents leads to sets of linear algebraic equations relatively to $N_{Au}$, $N_{Ad}$, $V_{Au}$, and
$V_{Ad}$. Condition of existence of nonzero solutions for amplitudes of perturbations gives us a dispersion equation.

\begin{figure}
\includegraphics[width=8cm,angle=0]{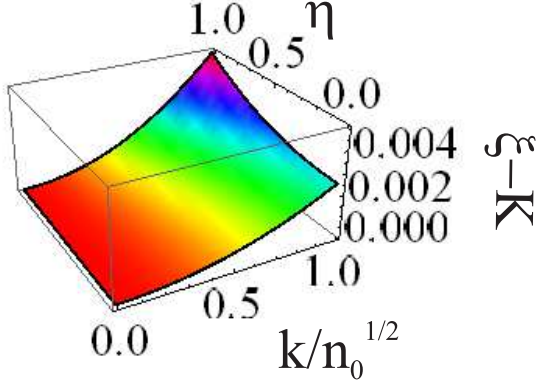}
\caption{\label{SUSD2D disp plane 1} (Color online) The figure shows dispersion of the Langmuir wave in plane-like 2DEGs and its dependence on spin-polarization. On this figure we use $\xi=\xi_{pl}$. Corresponding analytical solution is presented by formula (\ref{SUSD2D dispersion plane dimless}) with plus before the square root.}
\end{figure}
\begin{figure}
\includegraphics[width=8cm,angle=0]{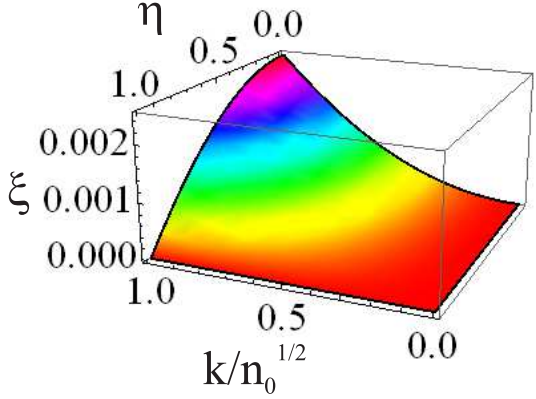}
\caption{\label{SUSD2D disp plane 2} (Color online) The figure shows dispersion and spin polarisation dependencies of the spin-electron acoustic wave for plane-like 2DEGs. On this figure we use $\xi=\xi_{pl}$. Corresponding analytical solution is presented by formula (\ref{SUSD2D dispersion plane dimless}) with minus before the square root.}
\end{figure}

Difference of spin-up and spin-down concentrations of electrons $\Delta n=n_{0\uparrow}-n_{0\downarrow}$ is caused  by external magnetic field. Since electrons are negative their spins get preferable direction opposite to the external magnetic field $\eta\equiv\frac{\Delta n}{n_{0}}=\tanh(\frac{\gamma_{e}B_{0}}{\varepsilon_{Fe,2D}})=-\tanh(\frac{\mid\gamma_{e}\mid B_{0}}{\varepsilon_{Fe,2D}})$, where $\varepsilon_{Fe,2D}=\pi n_{0}\hbar^{2}/m$, and $n_{0}=n_{0u}+n_{0d}$. 

We consider plasmas in the uniform constant external magnetic field. We see that in linear approach numbers of electrons of each species conserves.

Let us presents results for wave dispersion. We start with the plane-like 2DEG.

We assume that the external magnetic field is perpendicular to the plane, where the electron gas is located. Dispersion dependencies appears in the following form
$$\omega^{2}-\Omega^{2}=\frac{1}{2}\Biggl(\omega_{L,u}^{2}+\omega_{L,d}^{2}+(U_{u}^{2}+U_{d}^{2})k^{2}$$
$$\pm\biggl[(\omega_{L,u}^{2}+\omega_{L,d}^{2})^{2}+(U_{u}^{2}-U_{d}^{2})^{2}k^{4}$$
\begin{equation}\label{SUSD2D dispersion plane} +2 k^{2}(U_{u}^{2}-U_{d}^{2})(\omega_{L,u}^{2}-\omega_{L,d}^{2})\biggr]^{1/2}\Biggr),\end{equation}
where
$\omega_{L,s}^{2}=2\pi e^{2}n_{0,s}k/m$ is the two dimensional Langmuir frequency for species $s$ of electrons located in a plane, $\omega_{L}^{2}=\omega_{L,u}^{2}+\omega_{L,d}^{2}$ is the full Langmuir frequency, $\Omega=q_{e}B_{0}/(mc)$ is the cyclotron frequency, $U_{s}^{2}=\frac{2\pi\hbar^{2}}{m^{2}}n_{0s}+\frac{\hbar^{2}k^{2}}{4m^{2}}$ presents combined contribution of the Fermi pressure and the quantum Bohm potential.

Dropping the quantum Bohm potential and passing into dimensionless variables we obtain
\begin{equation}\label{SUSD2D dispersion plane dimless} \xi_{pl}=\frac{1}{2}\biggl(K+\Lambda K^{2}\pm\sqrt{K^{2}+2\Lambda K^{3}\eta^{2}+\Lambda^{2}K^{4}\eta^{2}}\biggr),\end{equation}
where $\xi_{pl}=(\omega^{2}-\Omega^{2})/\omega_{ch}^{2}$, $K=k/\sqrt{n_{0}}$, $\Lambda=r_{B}\sqrt{n_{0}}$, with $\omega_{ch}^{2}=2\pi e^{2}n_{0}^{\frac{3}{2}}/m$, $r_{B}=\hbar^{2}/(me^{2})$ is the Bohr radius. Numerical analysis of formula (\ref{SUSD2D dispersion plane dimless}) is presented on Figs. (\ref{SUSD2D disp plane 1}) and (\ref{SUSD2D disp plane 2}).
The spin-electron acoustic wave (SEAW) exists at intermediate spin polarisation. It disappears at both the zero and full spin polarizations.

Fig. (\ref{SUSD2D disp plane 1}) shows the dimensionless shift of dispersion dependence of the Langmuir wave $\xi_{pl}-K$ from the Langmuir frequency square for 2DEG in plane $\omega_{L}^{2}$. We note that 2D Langmuir frequency $\omega_{L}^{2}$ is the linear function of the wave vector $k=K\sqrt{n_{0}}$. Fig. (\ref{SUSD2D disp plane 1}) depicts dependence of the shift $\xi_{pl}-K$ on the dimensionless wave vector $K$ and spin polarisation $\eta$. We see that $\xi_{pl}-K$, as well as $\xi_{pl}$, increases with the increase of the wave vector. It happens due to the pressure of degenerate electron gas. Fig. (\ref{SUSD2D disp plane 1}) also shows that the growth of $\xi_{pl}(K)$ increases with increasing of the spin-polarisation. This effect appears due to dependence of pressure contribution on spin polarization via different occupation of quantum states by spin-up and spin-down electrons. We consider this effect by different pressure of spin-up and spin-down electrons. Nevertheless this effect can be included in single fluid model of electrons via corresponding equation of state (see Ref. \cite{Andreev 1403 exchange} for more details).

Dispersion dependence of the SEAW $\xi_{pl}(K)$ at different spin polarisation $\eta$ is depicted on Fig. (\ref{SUSD2D disp plane 2}). We see that  frequency square of the SEAW $\xi_{pl}(K)$ on three orders smaller than the Langmuir frequency. We also see that frequency of the SEAW increases with the increase of the wave vector $k$. The rate of the increasing slows down with growth of the spin polarisation.

\begin{figure}
\includegraphics[width=8cm,angle=0]{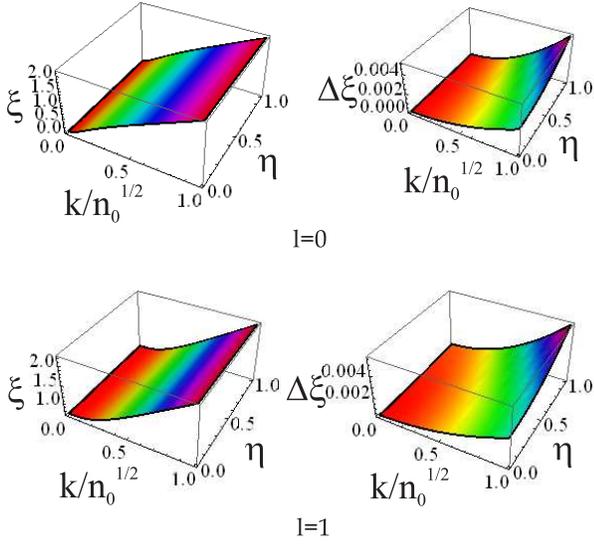}
\caption{\label{SUSD2D disp cyl Lang 1} (Color online) The figure shows dispersion of the Langmuir waves on cylindrical surface $\xi=\xi_{cyl}$ and shift of frequency square of the Langmuir waves $\Delta\xi$ from the Langmuir frequency square $\omega_{L,cyl}^{2}$ for $l=0$ and $l=1$. Analytical expression of dispersion is given by formula (\ref{SUSD2D dispersion cylinder dimless}) with the plus before the square root.}
\end{figure}
\begin{figure}
\includegraphics[width=8cm,angle=0]{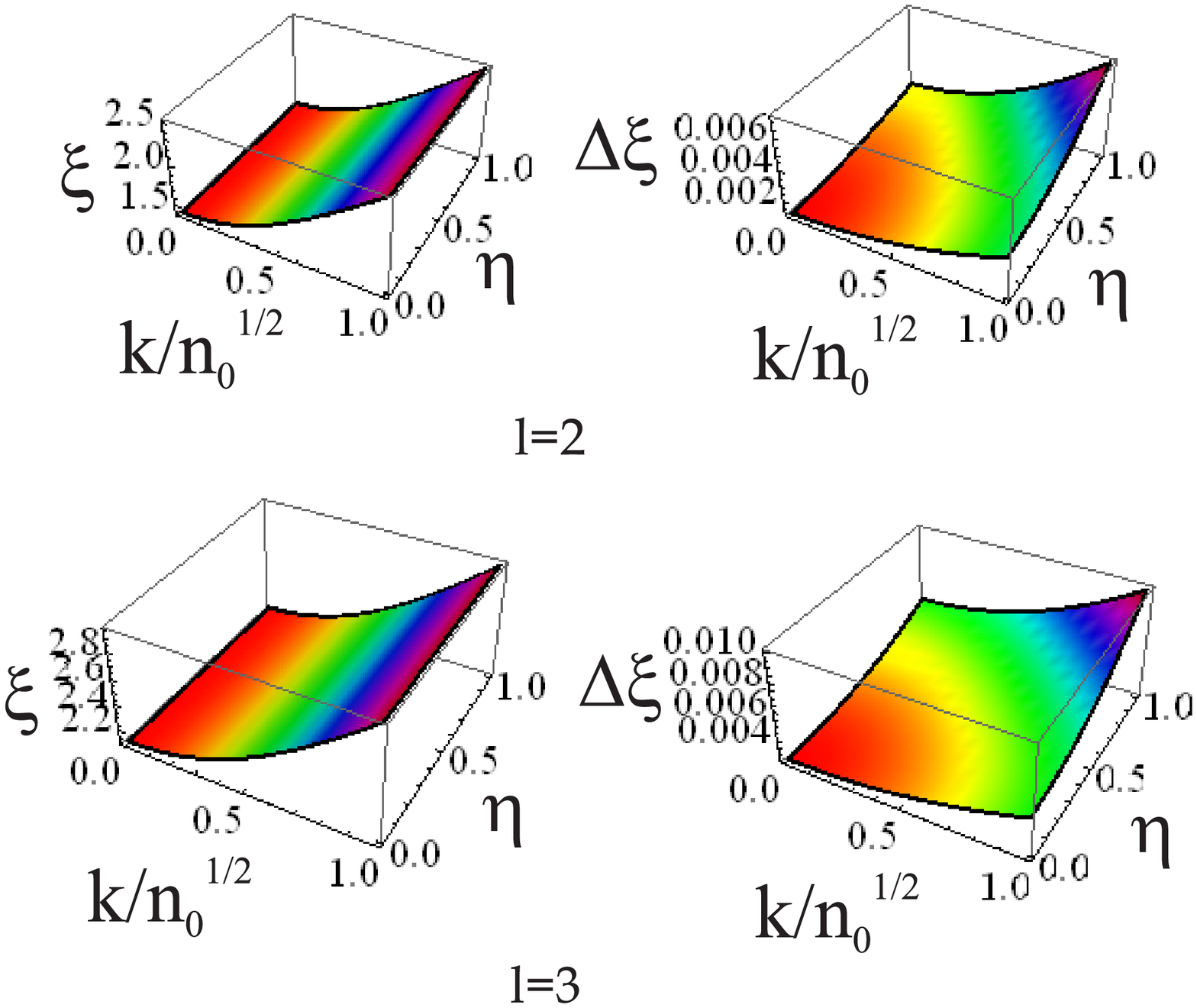}
\caption{\label{SUSD2D disp cyl Lang 2} (Color online) The figure shows dispersion of the Langmuir waves on cylindrical surface $\xi=\xi_{cyl}$ and shift of frequency square of the Langmuir waves $\Delta\xi$ from the Langmuir frequency square $\omega_{L,cyl}^{2}$ for $l=2$ and $l=3$. Analytical expression of dispersion is given by formula (\ref{SUSD2D dispersion cylinder dimless}) with the plus before the square root.}
\end{figure}
\begin{figure}
\includegraphics[width=8cm,angle=0]{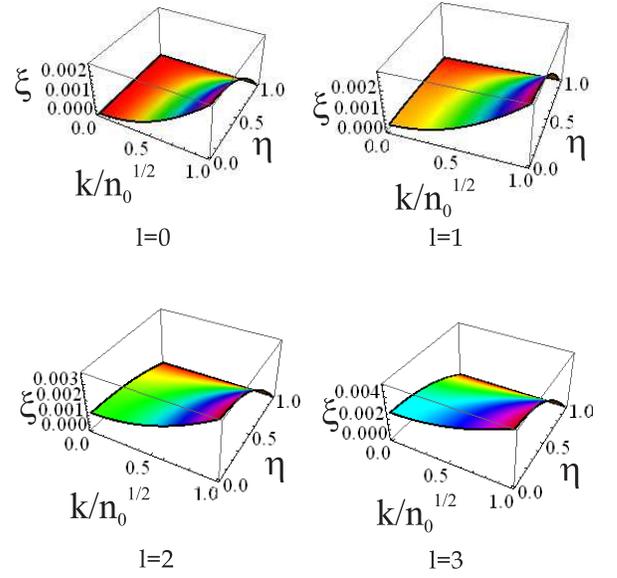}
\caption{\label{SUSD2D disp cyl SEAW} (Color online) The figure presents spectrums of the spin-electron acoustic wave at different values of discrete wave number $k_{\varphi}=l/R$. We present dispersion surfaces for $l=$0, 1, 2, 3. Analytical expression of dispersion is given by formula (\ref{SUSD2D dispersion cylinder dimless}) with the minus before the square root.}
\end{figure}

We present now results for 2DEG on the cylindric surface.

External magnetic field is parallel to the axis of the cylinder (nanotube), where electron gas is located. Corresponding dispersion dependence appears as
$$\omega^{2} =\frac{1}{2}\Biggl(\omega_{L,cyl}^{2}+\frac{l^{2}}{R^{2}}\biggl[v_{u}^{2}+v_{d}^{2}+\frac{\hbar^{2}}{2m^{2}}\biggl(k^{2}+\frac{l^{2}+2}{R^{2}}\biggr)\biggr]$$
$$+k^{2}\biggl[v_{u}^{2}+v_{d}^{2}+\frac{\hbar^{2}}{2m^{2}}\biggl(k^{2}+\frac{l^{2}}{R^{2}}\biggr)\biggr]$$
$$\pm\biggl[\biggl(k^{2}+\frac{l^{2}}{R^{2}}\biggr)^{2}[v_{u}^{2}-v_{d}^{2}+e^{2}RG(n_{0u}-n_{0d})/m]^{2}$$
\begin{equation}\label{SUSD2D dispersion cylinder} +4\omega_{L,cyl,u}^{2}\omega_{L,cyl,d}^{2}\biggr]^{1/2}\Biggr),\end{equation}
where
\begin{equation}\label{SUSD2D Lengm freq on Cyl} \omega_{L,cyl,s}^{2}=\frac{G e^{2}n_{0,s}}{m}R\biggl(k^{2}+\frac{l^{2}}{R^{2}}\biggr),\end{equation}
is the Langmuir frequency of electron gas on the cylinder for species $s=\{u=\uparrow, d=\downarrow\}$ of electrons, $\omega_{L,cyl}^{2}=\omega_{L,cyl,u}^{2}+\omega_{L,cyl,d}^{2}$ is the full Langmuir frequency, $v_{s}^{2}=2\pi\hbar^{2}n_{0s}/m^{2}$ presents contribution of the Fermi pressure, and $G=G(R,k,l)=4\pi I_{l}(kR)K_{l}(kR)$,
with $I_{l}(x)$, $K_{l}(x)$ are the modified Bessel functions.

Dimensionless form of dispersion dependence (\ref{SUSD2D dispersion cylinder}) can be written as
$$\xi_{cyl}=\frac{1}{2}\biggl(K^{2}+l^{2}Y^{2}\biggr)\Biggl(\frac{G}{2\pi Y}+\Lambda$$
\begin{equation}\label{SUSD2D dispersion cylinder dimless} \pm\sqrt{\biggl(\frac{G}{2\pi Y}\biggr)^{2}+2\eta^{2}\frac{G}{2\pi Y}\Lambda+\eta^{2}\Lambda^{2}}\Biggr),\end{equation}
where $\xi_{cyl}=\omega^{2}/\omega_{ch}^{2}$, $K=k/\sqrt{n_{0}}$, $\Lambda=r_{B}\sqrt{n_{0}}$, $Y=1/(R\sqrt{n_{0}})$.

%$R=30$ nm.

Figs. (\ref{SUSD2D disp cyl Lang 1}) and (\ref{SUSD2D disp cyl Lang 2}) show behavior of the Langmuir wave dispersion in 2DEG on the cylindric surface. At numerical analysis we assume that radius of the cylinder equal 30 nm. At $l=0$ frequency square of the Langmuir wave $\xi$ almost linearly depends on the wave vector $k$, but frequency shift $\Delta\xi=\xi-\omega_{Le,cyl}^{2}(k,l)/\omega_{ch}^{2}$ shows small difference from the linear growth. This difference is related to the pressure of degenerate gas. We also see that the pressure contribution dependes on spin polarisation $\eta$. At $l\neq0$, $\xi(k)$ reveals almost parabolic dependence due to structure of the Langmuir frequency square for electron gas on the cylindric surface.

Increase of the discrete wave number $k_{\varphi}=l/R$ gives an increase of whole dispersion surface. At the same time, the increase of $l$ leads to increase of the shift $\Delta\xi(K,\eta)$. Form of surfaces describing the shift $\Delta\xi(K,\eta)$ also changes with increasing of $l$: area of small wave numbers and large spin polarisation grows up faster than other areas.

Fig. (\ref{SUSD2D disp cyl SEAW}) describes the SEAW in the cylindric 2DEG for $l$=0, 1, 2, 3. General behavior shows resemblance to dispersion of the SEAW in plane-like 2DEG: increase of $\xi(k)$, lowering of $\xi(k)$ with increasing of spin polarisation $\eta$. Increasing of $l$ reveals in modification of dispersion surface of the SEAW. Values of $\xi$ increase with the increase of $l$. Moreover, the area of small wave vectors and small spin polarisation increases relatively faster. Modifications of $\xi_{cyl}$ for the SEAW and $\Delta\xi_{cyl}$ for the Langmuir wave happening with the change of $l$ are different. Different areas of these surfaces show relative growth. In both cases these areas are located at small wave vectors. However they are located at different values of the spin polarisation $\eta$. Relative increase of $\Delta\xi_{cyl}$ for the Langmuir wave is at large $\eta$. Whereas, the relative increase of $\xi_{cyl}$ for the SEAW is at small spin polarisation $\eta$.

In this paper we have discussed increase of frequency of the 2D Langmuir excitations due to different occupation of spin-up and spin-down states in the electron gas located in the external magnetic field. We have demonstrated existence of the SEAW in 2D structures (planes and nanotubes). We have described properties of the SEAW. These results have been obtained by means of the SSE-QHD.

\end{document}